\documentclass[conference]{IEEEtran}
\usepackage{multirow}
\IEEEoverridecommandlockouts
\usepackage{cite}
\usepackage{amsmath,amssymb,amsfonts}
\usepackage{algorithmic}
\usepackage{graphicx}
\usepackage{textcomp}
\usepackage{xcolor}
\def\BibTeX{{\rm B\kern-.05em{\sc i\kern-.025em b}\kern-.08em
    T\kern-.1667em\lower.7ex\hbox{E}\kern-.125emX}}

\begin{document}
\title{Predicting Cognitive Decline: A Multimodal AI Approach to Dementia Screening from Speech\\
}
\author{
    \IEEEauthorblockN{
        Lei Chi\IEEEauthorrefmark{1},
        Arav Sharma\IEEEauthorrefmark{1}, Ari Gebhardt\IEEEauthorrefmark{1}, and
        Joseph Colonel\IEEEauthorrefmark{2}}
    \IEEEauthorblockA{
        \IEEEauthorrefmark{1}Department of Electrical Engineering, The Cooper Union for the Advancement of Science and Art, New York, US\\
        \IEEEauthorrefmark{2}Icahn School of Medicine at Mount Sinai, New York, US\\
        lei.chi@cooper.edu, arav.sharma@cooper.edu, ari.gebhardt@cooper.edu, joseph.colonel@mssm.edu}
}
\maketitle

\begin{abstract}
Recent progress has been made in detecting early stage dementia entirely through recordings of patient speech. Multimodal speech analysis methods were applied to the PROCESS challenge, which requires participants to use audio recordings of clinical interviews to predict patients as healthy control, mild cognitive impairment (MCI), or dementia and regress the patient's Mini-Mental State Exam (MMSE) scores. The approach implemented in this work combines acoustic features (eGeMAPS and Prosody) with embeddings from Whisper and RoBERTa models, achieving competitive results in both regression (RMSE: 2.7666) and classification (Macro-F1 score: 0.5774) tasks. Additionally, a novel two-tiered classification setup is utilized to better differentiate between MCI and dementia. Our approach achieved strong results on the test set, ranking seventh on regression and eleventh on classification out of thirty-seven teams, exceeding the baseline results.
\end{abstract}


\section{Introduction}
Dementia is a global healthcare challenge marked by a incremental cognitive decline that substantially affects memory, reasoning, and problem-solving skills \cite{knopman2003essentials}. As a neurodegenerative condition, dementia manifests via the decay of cognitive functions, affecting language, perception, and motor functions \cite{thabtah2020correlation}. A 2019 study indicates that approximately 50 million people worldwide live with dementia, with this number projected to triple by 2050 \cite{ADI2024}. Similarly, Mild Cognitive Impairment (MCI) represents an intermediate stage between normal cognitive aging and dementia, characterized by mild cognitive dysfunction without significant functional impairment. Alarmingly, dementia remains significantly underdiagnosed globally, with an estimated 75\% of cases going undetected \cite{Gauthier2021}. This diagnostic gap, coupled with the aging global population, has created a need for innovative solutions to facilitate remote diagnosis and improve healthcare accessibility.

Early detection of dementia is vital as it enables rapid intervention to form treatment strategies that can help manage symptoms and slow disease progression \cite{Livingston2017}. However, accurate early diagnosis has serious difficulties, requiring comprehensive clinical assessment that includes detailed patient history, cognitive evaluation, and careful elimination of other potential medical and psychiatric conditions that could manifest similar symptoms \cite{Johnson2021}. The complexity of this diagnostic process, along with limited healthcare resources, often results in delayed or missed diagnoses, particularly in regions with limited access to specialist care.
Recent advances in Artificial Intelligence (AI) and Machine Learning (ML) technologies provide opportunities for innovative solutions to these diagnostic challenges \cite{ADI2024}. Although traditional diagnostic methods are highly dependent on specialist assessment, data-driven approaches can provide accessible and cost-effective screening tools to support healthcare providers in early detection and accurate diagnosis, especially in primary care settings where most initial assessments occur. In particular, a 2024 study demonstrated that AI-based speech analysis could predict the progression from MCI to Alzheimer's disease with a precision greater than 78\% \cite{Huang2024}, showing the potential of these technologies in revolutionizing dementia diagnosis.

Speech analysis has emerged as a proven method for the early detection of cognitive decline. Connected speech analysis, which examines various aspects of spontaneous speech, has shown high sensitivity in detecting language impairments associated with MCI and early-stage dementia\cite{Mueller2018}. The method presented in this paper focuses on prosodic features such as speech rate, hesitation patterns, and changes in fundamental frequency and formants, which are indicative of cognitive decline\cite{Themistocleous2020}. For instance, individuals with MCI typically exhibit lower speech rates, longer hesitations, and alterations in pitch and voice quality\cite{Toth2018}.

Our research addresses the critical need for early detection through the development of an advanced speech analysis system that combines state-of-the-art deep learning models with traditional acoustic features. We employ a multi-task approach, incorporating various speech tasks including picture description and lexical-semantic retrieval. This combination of tasks has been shown to outperform single-task approaches in detecting cognitive impairment \cite{Fraser2019}. Our system analyzes both linguistic content and acoustic characteristics, with particular attention to prosodic elements such as pause patterns and postpause speech, which can be early indicators of MCI \cite{Roark2011}.

This work seeks to advance a non-invasive, accessible, and accurate tool for early detection of cognitive decline. Dementia and MCI can then be detected earlier and managed more effectively, potentially transforming the landscape of dementia care and improving the quality of life of susceptible and diagnosed individuals.

\begin{figure}[htbp]
\centerline{\includegraphics[scale=0.5]{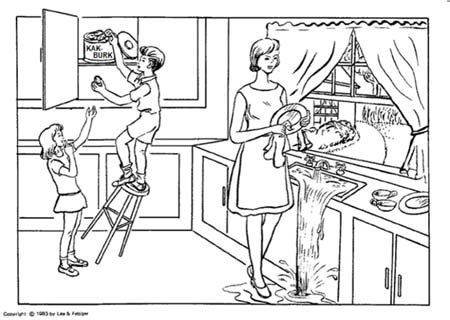}}
\caption{Cookie theft test prompt image}
\label{fig}
\end{figure}

\section{Background}

\subsection{Cognitive Assessment Tasks}

Three different tasks, each intended to assess a different facet of cognitive function, are incorporated\cite{tao2024early}. In the first task, Semantic Fluency, participants have one minute to name as many animals as they can\cite{205bc1c653a04def911237057e927572}. This test is essential for assessing name skills, linguistic proficiency, and semantic memory access, all of which are frequently impacted in the early phases of cognitive decline. The task's efficacy stems from its capacity to evaluate linguistic expression and comprehension at the same time.

Participants in the second assignment, Phonemic Fluency, have one minute to come up with words that start with the letter P, eliminating proper nouns such as the names of persons or places\cite{205bc1c653a04def911237057e927572}. This task focuses on language processing executive functions and verbal fluency. The task establishes a controlled environment that evaluates word retrieval ability and cognitive flexibility under particular limits by limiting the first letter and eliminating specified categories.

\begin{figure}[h]
\centerline{\includegraphics[scale=0.5]{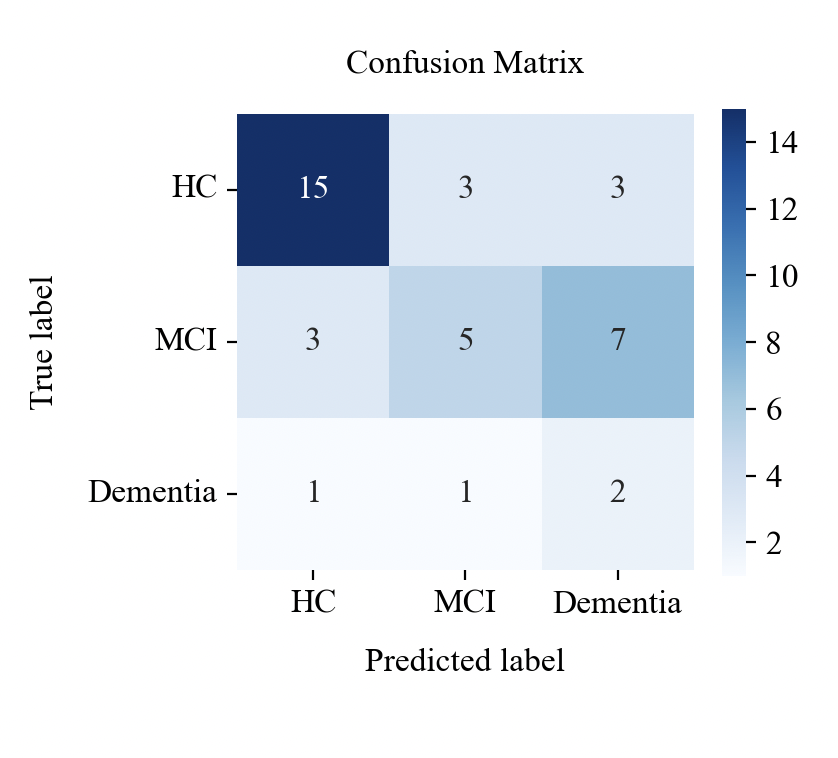}}
\caption{Confusion matrix of initial model}
\label{fig}
\end{figure}

The Cookie Theft picture description task, the third evaluation, is a commonly used technique in dementia detection studies\cite{goodglass2001bdae}. After viewing a standardized image of a kitchen scene, participants are asked to explain the details and activities they see. This extensive task simultaneously activates several cognitive domains, such as memory integration, narrative construction, and visual processing. The description's free-form format enables researchers to examine both planned and unplanned facets of language production.

\subsection{Mini-Mental State Examination (MMSE)}

The Mini-Mental State Examination (MMSE) is a commonly used cognitive screening tool designed to assess overall mental function. It consists of a 30-point questionnaire evaluating key cognitive domains, including orientation, attention, memory, language, and visual-spatial skills\cite{folstein1975mini}. The MMSE is widely used in clinical and research settings to screen for cognitive impairment and track changes over time, particularly in conditions like Mild Cognitive Impairment (MCI) and dementia.

\subsection{Audio Feature Sets}

We used audio feature sets to extract information from the recordings. These audio sets were designed to extract information relevant to different aspects of the recording while being limited enough in scope to be easily interpreted by the machine learning methods.

The eGeMAPS feature set\cite{Eyben2016} was designed to extract information from voice for a wide range of applications. eGeMAPS is commonly used as a baseline for voice analysis tasks. The feature set includes pitch, jitter, loudness, and spectral parameters, among others. 

Prosody features\cite{dehak2007modeling}, as implemented in the DisVoice package\cite{vasquez2018towards}, is a more targeted feature set than eGeMAPS designed specifically for prosodic aspects of voice, which hold special relevance to voice degeneration resulting from mental decline. This includes information about the duration and quality of voiced and unvoiced speech as well as pauses in speech.

\subsection{Deep Learning Models}

Deep learning models were used to extract and condense the pertinent information from the patient recordings. In contrast to traditionally extracted features, deep learning features contain more information about the recording but are less understandable by human observers. The deep learning systems introduce additional acoustic and linguistic information.

Wav2Vec2 \cite{baevski2020wav2vec} is a deep learning software developed by Meta which condenses audio into a sequence of embeddings, originally designed for audio transcription. These embedding sequences contain the most pertinent information from the audio recordings and as a result are easier to process by machine learning systems than raw audio.

Whisper \cite{Radford2023} is a speech transcription software designed by OpenAI. Whisper employs an encoder-decoder transformer architecture, where the encoder generates a sequence of embeddings, much like Wav2Vec2, and the decoder uses the embeddings to generate transcriptions. We use the Whisper encoder embeddings in the same manner as the Wav2Vec2 embeddings.

To process the semantic and grammatical aspects of patient speech, we transcribe the audio using Whisper and then process those transcriptions using the RoBERTa system \cite{Liu2019}. RoBERTa is a natural language processing system which can take in text and reduce the input down to one embedding.

\begin{figure*}[t!]
    \centering
    \includegraphics[width=\textwidth]{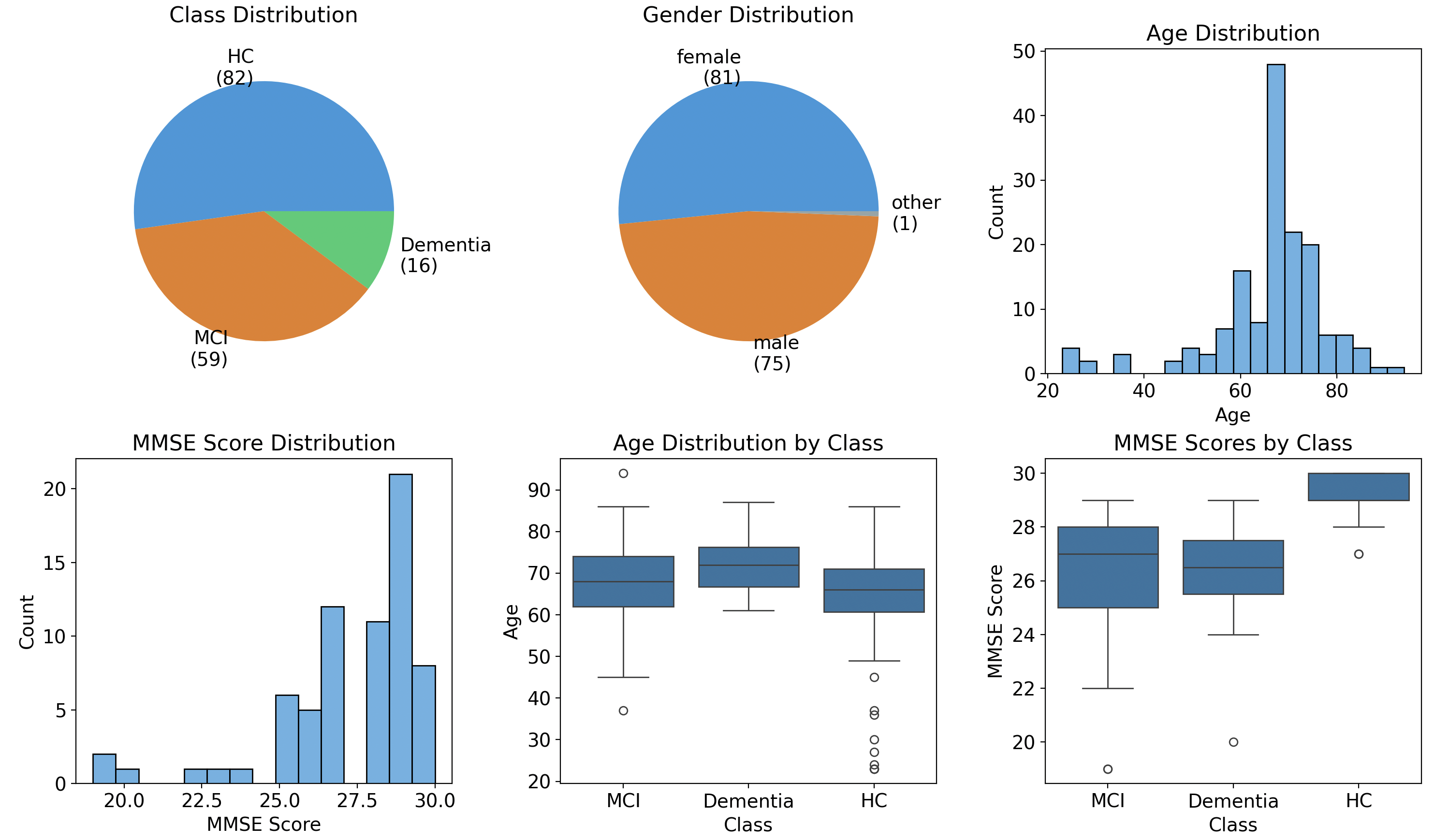} 
    \caption{PROCESS dataset demographic metrics}
    \label{fig:two_column_image}
\end{figure*}

\subsection{Performance Metrics}
The evaluation framework incorporates both classification and regression metrics to comprehensively assess the system's performance. Below, these metrics are detailed, starting with basic classification measures and their macro-level extensions, followed by regression metrics and our combined scoring approach.
\subsubsection{Classification Metrics}
At the instance level, three fundamental metrics are employed. Precision (Eq. 1) measures the ratio of accurate positive predictions to all positive predictions, crucial for minimizing false positives that could lead to unnecessary medical procedures. Recall (Eq. 2) indicates the proportion of actual positive cases correctly identified, essential for preventing missed diagnoses in healthcare settings. The F1 Score (Eq. 3) combines these metrics to provide a balanced measure of model performance.
\subsubsection{Macro-level Extensions}
For imbalanced datasets, macro-level metrics that give equal weight to each diagnostic class are used. Macro-Precision (Eq. 4) averages precision across all categories, while Macro-Recall (Eq. 5) averages recall scores. The Macro-F1 Score (Eq. 6) provides a balanced evaluation across all classes, preventing bias toward more frequent diagnostic categories.
\subsubsection{Regression and Combined Metrics}
For quantitative predictions of MMSE scores, Root Mean Squared Error (RMSE) (Eq. 7) is used, where lower values indicate more accurate predictions. Following the PROCESS 2025 Challenge evaluation criteria, we introduce a combined score (Eq. 8) that integrates both classification and regression performance for comprehensive ranking of participants. This scoring system, which balances F1 Score and RMSE performance, determines the final competition rankings.

The complete mathematical formulation of the performance metrics is as follows:

\begin{equation}
\text{Precision}_i = \frac{\text{TP}_i}{\text{TP}_i + \text{FP}_i}
\end{equation}

\begin{equation}
\text{Recall}_i = \frac{\text{TP}_i}{\text{TP}_i + \text{FN}_i}
\end{equation}

\begin{equation}
\text{F1}_i = 2 \times \frac{\text{Precision}_i \times \text{Recall}_i}{\text{Precision}_i + \text{Recall}_i}
\end{equation}

\begin{equation}
\text{Macro-Precision} = \frac{1}{N} \times \sum_{i=1}^N \text{Precision}_i
\end{equation}

\begin{equation}
\text{Macro-Recall} = \frac{1}{N} \times \sum_{i=1}^N \text{Recall}_i
\end{equation}

\begin{equation}
\text{Macro-F1 Score} = \frac{1}{N} \times \sum_{i=1}^N \text{F1}_i
\end{equation}

\begin{equation}
\text{RMSE} = \sqrt{\frac{\sum_{i=1}^N(\hat{y}_i - y_i)^2}{N}}
\end{equation}

\begin{equation}
S_k = \frac{\text{F1Score}_k}{\sum_j^T \text{F1Score}_j} + 1 - \frac{\text{RMSE}_i}{\sum_j^T \text{RMSE}_j}
\end{equation}

\noindent where:
\begin{itemize}
\item $N$ is the number of diagnostic classes
\item $\text{TP}_i$, $\text{FP}_i$, $\text{FN}_i$ are true positives, false positives, and false negatives for the $i$-th class
\item $y_i$ is the actual MMSE score and $\hat{y}_i$ is the predicted MMSE score
\item $S_k$ is the total score of participant $k$
\item $T$ is the total number of participants
\end{itemize}

\subsection{Dataset}

The PROCESS Challenge dataset includes data from 157 subjects, split between training and validation. The demographic metrics of the dataset are depicted in Fig. 3. In order to measure different facets of verbal communication and cognitive function, each participant took part in the Cookie Theft Description task, the Phonemic Fluency Test, and the Semantic Fluency Test. These exercises offer a thorough framework for examining dementia-related speech patterns. The dataset was highly imbalanced, with a large number of healthy control and a small number of dementia cases.

\section{Related Work}

Research on speech-based dementia detection has expanded significantly, leveraging linguistic, acoustic, and deep learning-based approaches. This section reviews key studies, benchmarking challenges, and their influence on our proposed approach.

\subsection{Speech-Based Dementia Detection}

Prior studies demonstrated that cognitive decline manifests in spontaneous speech through increased pauses, word-finding difficulties, and syntactic simplifications. Fraser \textit{et al.} achieved 81-82\% accuracy on the DementiaBank dataset using lexical and syntactic features \cite{fraser2015linguistic}. Weissenbacher \textit{et al.} further refined this approach, reaching 86\% accuracy by incorporating larger datasets \cite{weissenbacher-etal-2016-automatic}. Additionally, Yancheva \textit{et al.} showed that speech features could predict MMSE scores, highlighting their role in tracking cognitive decline \cite{yancheva2015using}.

\subsection{ADReSS and ADReSSo Challenges}

To standardize research in this field, the Alzheimer’s Dementia Recognition through Spontaneous Speech (ADReSS) challenge (INTERSPEECH 2020) introduced a balanced dataset for AD classification and MMSE score regression \cite{luz2021alzheimer}. This dataset mitigated demographic biases and enabled systematic comparisons. The ADReSSo challenge (2021) extended this work by focusing solely on raw speech, requiring models to function without manual transcripts \cite{luz2021detecting}. Winning models from ADReSS achieved 85-89.6\% classification accuracy, demonstrating that combining linguistic and acoustic features yields the best results \cite{balagopalan2020bert}.

\begin{figure*}[t!]
    \centering
    \includegraphics[width=\textwidth]{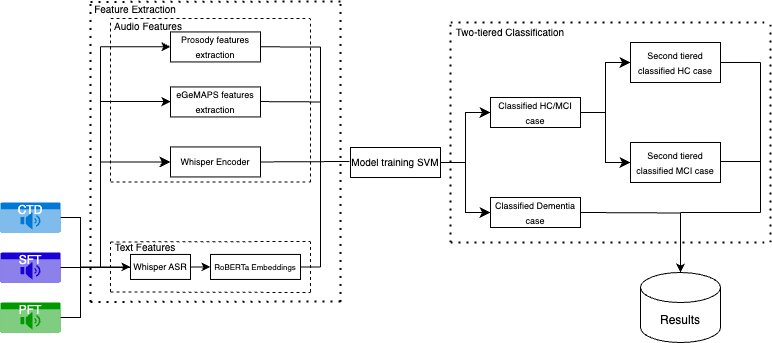} 
    \caption{Two-tiered classification architecture.}
    \label{fig:two_column_image}
\end{figure*}

\subsection{Comparison of Methodologies}

Approaches to dementia detection fall into three broad categories, which are:

\begin{itemize}
    \item Linguistic Approaches – Features are extracted from transcribed speech, analyzing word choice, fluency, and syntactic complexity. Transformer-based language models (e.g., BERT) have improved classification accuracy in recent challenges \cite{balagopalan2020bert}.
    \item Acoustic Approaches – Features related to prosody, articulation, and voice quality are extracted, as in the eGeMAPS the DisVoice Prosody feature sets. Traditional classifiers achieve 70-80\% accuracy, but perform worse than text-based models \cite{8910399}.
    \item Multimodal Approaches – Use wav2vec2 or CNNs to learn from raw audio. Recent methods integrate both linguistic and acoustic features using multimodal architectures, achieving state-of-the-art results \cite{syed2020automated}.
\end{itemize}

\subsection{Relation to our System}

Our approach builds upon these findings by integrating acoustic and linguistic embeddings using a deep learning-based framework. Unlike traditional methods that analyze transcripts and speech separately, our method employs an end-to-end model that jointly processes both modalities. This allows us to achieve higher classification accuracy and a lower MMSE prediction error, outperforming previous ADReSS baselines. Additionally, our two-tier classification system enhances its ability to distinguish between MCI and dementia, an area where past models have struggled. The next section provides technical details of our methodology.

\section{Methods}

\subsection{Feature Extraction}

In order to capture both the semantic and acoustic aspects of speech, our method combines deep learning models with traditional feature sets. RoBERTa is used to extract linguistic embeddings from the transcriptions generated by Whisper. The mean and standard deviation of the embedding sequences generated by Whisper's acoustic front-end feature extractor were also used. Conventional features were also added to these, such as the eGeMAPS and prosodic feature sets.

\subsection{Two-Tiered Classification System}

Previous challenges in cognitive impairment detection typically focused on the binary classification between healthy control (HC) and dementia. The PROCESS challenge increases complexity by requiring simultaneous detection of HC, mild cognitive impairment (MCI), and dementia. The distinction between dementia and MCI is particularly challenging because these conditions represent points on a continuous spectrum of cognitive decline, with MCI often being a transitional stage between normal cognition and dementia. The symptoms and cognitive markers can overlap significantly, with the primary differences appearing in severity rather than clear-cut distinguishing features. To address this complexity, a two-tiered classification approach using machine learning methods as shown in Fig. 4. The first classifier distinguishes dementia from non-dementia cases, while the second classifier separates healthy controls from impaired cases. The final classification is determined by combining these binary decisions. For instance, a sample classified as non-dementia by the first classifier and non-healthy by the second classifier would be labeled as MCI.

\begin{figure*}[t!]
    \centering
    \includegraphics[width=\textwidth]{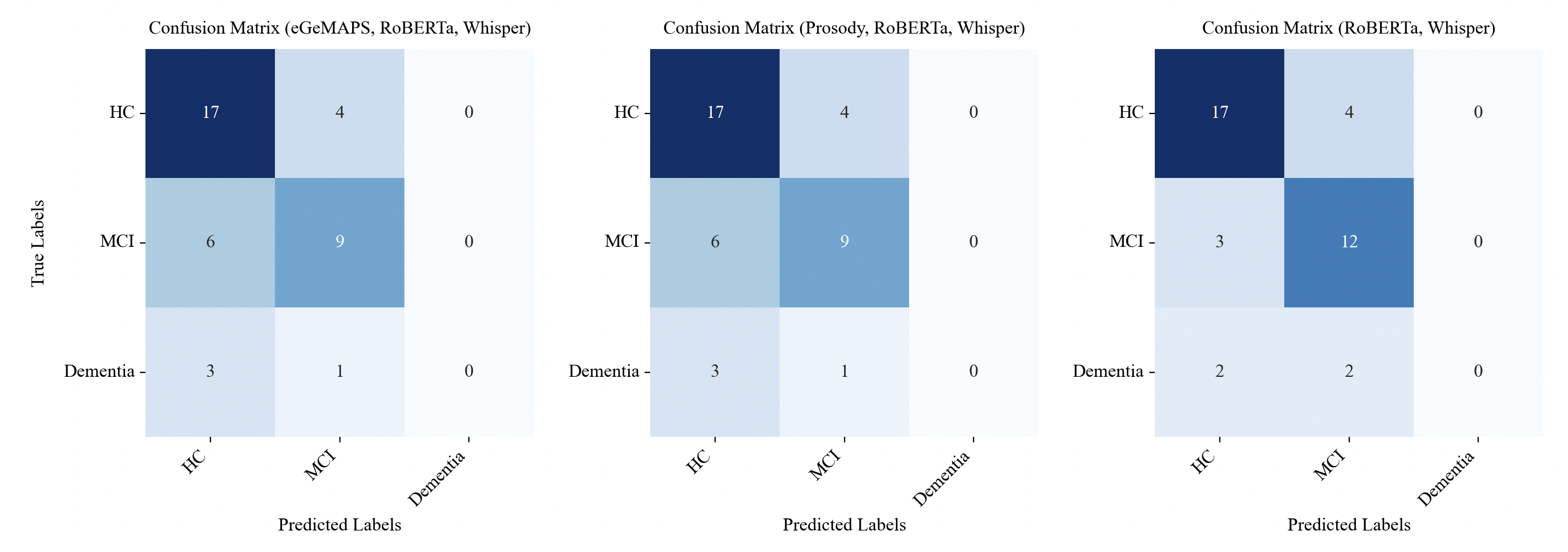}
    \caption{Confusion matrices for our three submitted models on the test set. From left to right: Base model (Whisper + RoBERTa embeddings), Model 2 (Base + eGeMAPS), and Model 3 (Base + prosody features). All models use the two-tiered SVM classification approach.}
    \label{fig:confusion_matrices}
\end{figure*}

\subsection{Method 1: CNN-XGBoost}

The first method employed a sequential classification strategy using Wav2Vec2 features extracted from segmented audio recordings. The first stage was composed a 1D Convolutional Neural Network (CNN) with clustering and time-weighted classification aggregation, in which the individual classifications of later clusters were given a greater influence in the final classification. The remaining samples were then classified as either MCI or HC using XGBoost. This forked classification structure demonstrated strong initial performance in diagnosing dementia as shown in Fig. 2. 

\subsection{Method 2: SVM}

Subsequent experimentation revealed that the Wav2Vec2 features introduced unnecessary model complexity, with performance actually improving when using only Whisper features. Furthermore, it was observed that deep learning models (CNNs and MLPs) were susceptible to overfitting on the dataset, considering the small training set size. Method 2 instead uses Support Vector Machines (SVMs), which provided better generalization and therefore higher performance on the test set. The performance metrics of our previous model (CNN-XGBoost) and the final model (SVM) are shown in Table \ref{tab:comparison}.

\begin{table}[!t]
\centering
\caption{Performance Comparison of Two-Tier Classification Approaches}
\label{tab:comparison}
\begin{tabular}{l@{\hspace{2em}}c@{\hspace{2em}}c}
\hline
 & CNN-XGBoost & SVM \\
& (Wav2Vec2) & (Whisper-RoBERTa) \\
\hline
Macro-Precision & 50.39\% & 48.00\% \\
Macro-Recall & 51.59\% & 53.70\% \\
Macro-F1 & 47.22\% & 50.60\% \\
\hline
\end{tabular}
\end{table}

\subsection{Regression Model}

 All retrieved embeddings and acoustic characteristics were integrated and a regression model based on Support Vector Regression (SVR) to predict MMSE scores. The features are consistent with those used in our classification method. This all-encompassing method makes it possible to combine many feature sets to improve the precision and resilience of dementia diagnosis.

\subsection{Implementation Details}

We created three model variations to investigate various feature combinations and their classification and regression performance. Using their advantages in text and speech processing, the basic model incorporates embeddings from Whisper and RoBERTa. Model 2 expands on the foundational model by adding eGeMAPS features, which record low-level descriptors associated with emotion and voice quality. By including prosodic features, Model 3 improves performance further. We used SMOTE (Synthetic Minority Over-sampling Technique)\cite{chawla2002smote} to alleviate class imbalance and provide balanced training data. In order to verify model performance and avoid overfitting, we used 5-fold cross-validation throughout development and grid search for hyperparameter adjustment.

\section{Results and Discussion}

Our evaluation encompasses both classification and regression tasks using standard metrics for cognitive decline detection. Table \ref{tab:results} summarizes the performance across all model variants, including baseline approaches.

\begin{table}[ht]
\centering
\caption{Performance Results on PROCESS Challenge Test Set}
\label{tab:results}
\begin{tabular}{lccccc}
\hline
\textbf{Model Variant} & \multicolumn{2}{c}{\textbf{Classification}} & \multicolumn{2}{c}{\textbf{Regression}} \\
\cline{2-5}
& Macro-F1 Score & Rank & RMSE & Rank \\
\hline
\multicolumn{5}{l}{\textbf{Baseline Models}} \\
eGeMAPS (SVC/SVR)$^{1}$ & 0.5500 & 21 & 4.4000 & 76 \\
eGeMAPS (RFC/RFR)$^{2}$ & 0.5330 & 25 & 3.1700 & 57 \\
RoBERTa$^{3}$ & 0.3289 & 95 & 2.9850 & 37 \\
\hline
\multicolumn{5}{l}{\textbf{Our Models}} \\
Base Model$^{4}$ & 0.4306 & 72 & 2.7666 & 9 \\
+ eGeMAPS$^{5}$ & 0.5774 & 16 & 2.7821 & 11 \\
+ Prosody$^{6}$ & 0.5774 & 15 & 2.7844 & 12 \\
\hline
\multicolumn{5}{l}{\footnotesize{$^{1}$Support Vector Classification/Regression with eGeMAPS features}} \\
\multicolumn{5}{l}{\footnotesize{$^{2}$Random Forest Classification/Regression with eGeMAPS features}} \\
\multicolumn{5}{l}{\footnotesize{$^{3}$RoBERTa embeddings only}} \\
\multicolumn{5}{l}{\footnotesize{$^{4}$Whisper + RoBERTa embeddings with two-tier SVM}} \\
\multicolumn{5}{l}{\footnotesize{$^{5}$Base model + eGeMAPS acoustic features}} \\
\multicolumn{5}{l}{\footnotesize{$^{6}$Base model + prosody features}} \\
\end{tabular}
\end{table}

\subsection{Classification Results}

Our analysis began with evaluating baseline models to establish performance benchmarks. The eGeMAPS-based SVM classifier achieved a Macro-F1 score of 0.5500 (rank 21), while the Random Forest approach reached 0.5330 (rank 25). Notably, the RoBERTa-only baseline performed significantly lower with a Macro-F1 score of 0.3289 (rank 95), suggesting that linguistic features alone may be insufficient for reliable cognitive impairment detection.

Our top-performing model placed 15th overall with a Macro-F1 score of 0.5774, surpassing all baseline approaches. This improvement was achieved through acoustic feature fusion, particularly after incorporating eGeMAPS features into our base architecture. The addition of prosodic features maintained this high performance while slightly improving the model's rank to 15th place. The use of acoustic feature fusion resulted in a notable improvement in performance, underscoring the importance of merging several feature sets for reliable classification. Additionally, the model consistently performed accurately on all three speech tests, demonstrating its dependability and versatility in a range of linguistic circumstances. The confusion matrices for the three final models are shown in Fig. 5. 

\subsection{Regression Results}
For the regression task, MMSE scores were predicted to measure the severity of cognitive impairment. Our best model attained an RMSE of 2.7666, securing 9th place in the evaluation. This represents a substantial improvement over the baseline approaches, with the eGeMAPS SVR baseline achieving an RMSE of 4.4000 (rank 76) and the Random Forest regressor achieving 3.1700 (rank 57). Even the RoBERTa baseline, while performing better with an RMSE of 2.9850 (rank 37), still fell short of our model's performance. Notably, our base model exhibited superior regression performance compared to advanced variants, suggesting that simpler architectures may effectively capture the underlying patterns in MMSE data. Additionally, the model's performance remained stable across different combinations of features, reaffirming the robustness of the approach.

\section{Conclusion}

This study introduces a multimodal method that uses speech-based categorization and regression to identify cognitive impairment. Our system, which ranks seventh in regression and eleventh in classification in the PROCESS 2025 International Challenge, obtains competitive results in both tasks by combining classical acoustic characteristics with deep learning-based embeddings. A major obstacle to early identification is addressed by the suggested two-tiered classification system, which enhances the ability to distinguish between MCI and dementia. These findings demonstrate automated speech analysis's promise as a non-invasive, scalable method for early dementia identification.

There is significant room for future research to improve our system even further. We used a small and imbalanced dataset as part of the restrictions for the PROCESS challenge, which inherently limits the potential efficacy of our work. With a larger and more varied dataset, as can be found through DementiaBank\cite{lanzi2023dementiabank}, we would expect to have better performance even without any changes to our methodology. A larger dataset could also allow for training large deep learning systems on the dataset directly, which would overfit the PROCESS dataset. Additionally, to improve our systems efficacy as a tool for medical experts, we could make our system more transparent and human interpretable by researching what attributes of speech and language our system uses for prediction.

\section*{Acknowledgment}

The authors would like to graciously thank the PROCESS Challenge organizers for organizing and allowing us to participate in this challenge.

\bibliography{references}

\end{document}